\documentclass[prc,twocolumn,showpacs]{revtex4}

\usepackage{bm}
\usepackage{graphicx}
\usepackage{longtable}

\begin{document}

\title{The $^8$B Neutrino Spectrum}
\author{W. T. Winter}
\author{S. J. Freedman}
\affiliation{Nuclear Science Division, Lawrence Berkeley National Laboratory, California 94720}
\affiliation{Physics Department, University of California, Berkeley, California 94720}
\author{K. E. Rehm}
\author{J. P. Schiffer}
\affiliation{Physics Division, Argonne National Laboratory, Argonne, Illinois 60439}


\date{\today}

\begin{abstract}

Knowledge of the energy spectrum of $^8$B neutrinos is an important ingredient for interpreting experiments that detect energetic neutrinos from the Sun. The neutrino spectrum deviates from the allowed approximation because of the broad alpha-unstable $^8$Be final state and recoil order corrections to the beta decay. We have measured the total energy of the alpha particles emitted following the beta decay of $^8$B. The measured spectrum is inconsistent with some previous measurements, in particular with a recent experiment of comparable precision. The beta decay strength function for the transition from $^8$B to the accessible excitation energies in $^8$Be is fit to the alpha energy spectrum using the R-matrix approach. Both the positron and neutrino energy spectra, corrected for recoil order effects, are constructed from the strength function. The positron spectrum is in good agreement with a previous direct measurement. The neutrino spectrum disagrees with previous experiments, particularly for neutrino energies above 12 MeV.

\end{abstract}

\maketitle

\section{\label{sec:Intro}Introduction}

The most carefully studied component of the solar neutrino flux is due to neutrinos from the $\beta^+$ decay of $^8$B. The $^8$B neutrinos account for most of the signal in the Homestake $^{37}$Cl neutrino capture experiment \cite{HOM} and nearly all of the solar neutrino events in the Kamiokande \cite{KAM},  Super-Kamiokande \cite{SK}, and the Sudbury Neutrino Observatory (SNO) \cite{SNO} water-Cherenkov experiments. Results from the SNO heavy water detector demonstrate the existence of a $\nu_{\mu,\tau}$ component of the solar neutrino flux \cite{SNO}. The solar neutrino data is explained by flavor oscillations and non-zero neutrino mass \cite{OSC}. The recent results of the KamLAND reactor $\overline{\nu}_e$ disappearance experiment \cite{KAMLAND} support the oscillation interpretation \cite{OSC}. The neutrino oscillation solution implies that the solar $^8$B $\nu_e$ energy spectrum is distorted. Knowledge of the primary $^8$B neutrino spectrum is a necessary ingredient for the proper interpretation of the solar neutrino data.

\begin{figure}
\includegraphics[width=0.3\textwidth]{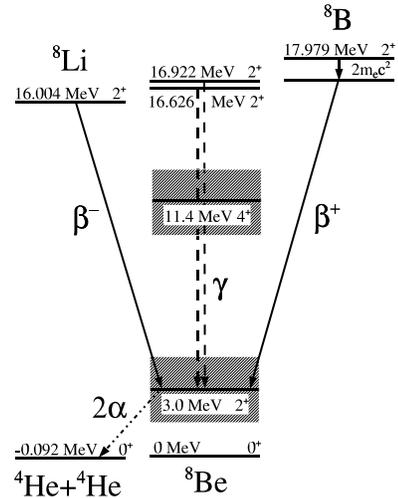}
\caption{\label{fig:levels} Nuclear levels in A=8 which lie below the $^8$B ground state. Spins, parities, and excitation energies relative to the $^8$Be ground state are indicated.}
\end{figure}

A diagram illustrating the $^8$B decay chain is shown in Fig. \ref{fig:levels}. The $^8$B ground state (J$^\pi$=2$^+$) undergoes an allowed $\beta^+$ transition to a broad range of excitation energies in the $\alpha$ unstable $^8$Be daughter. We define the $^8$B $\beta^+$ decay \textit{strength function} as the probability that a given excitation energy in $^8$Be will be populated. The strength function is determined by measurements of the $\alpha$ particle energy spectrum following the breakup of the daughter $^8$Be nucleus, and is necessary to construct the neutrino spectrum.

Transitions from $^8$B to the $^8$Be ground state (J$^\pi$=0$^+$) or the broad state at 11.4~MeV (J$^\pi$=4$^+$) are second forbidden and highly suppressed. States in $^8$Be with J$^\pi$=1$^+$,3$^+$ are not energetically accessible. The $^8$B $\beta^+$ decay thus proceeds exclusively through the resonance 2$^+$ structure in $^8$Be, described in the R-matrix formalism as a series of interfering 2$^+$ states. We have measured the total energy of the $\alpha$ particles emitted following $^8$B $\beta^+$ decay in a recent experiment \cite{WIN}, reviewed in Sec. \ref{sec:Alpha}. The data are analyzed in the framework of the many-level R-matrix approximation to fit the $\beta^+$ decay strength function, which is presented in a table and compared with the results of the previous precision measurement \cite{ORT} in Sec. \ref{sec:AlphaFits}. The R-matrix approach employed here is not essential for obtaining the neutrino spectrum, but provides a convenient way to characterize the experimental data.

The $^8$B neutrino spectrum is subject to corrections due to recoil order matrix elements. Measurements involving the $\beta^+$($\beta^-$) decay of $^8$B($^8$Li) and the radiative decays of the 2$^+$ doublet in $^8$Be with excitation energies near 16~MeV, shown in Fig. \ref{fig:levels}, are used to extract the recoil order matrix elements which contribute to $^8$B $\beta^+$ decay. A review of past recoil order measurements is presented in Sec. \ref{sec:Recoil}.

Both the $^8$B positron and neutrino energy spectra are deduced, using the strength function and applying recoil order and radiative corrections, in Sec. \ref{sec:Neutrino}. The agreement between this work and the previous direct measurement of the positron spectrum \cite{NFC} is discussed, and the neutrino spectrum is presented in a table.

\section{\label{sec:Alpha} The Alpha Spectrum Measurement}

A description of the $\alpha$-spectrum measurement discussed here has been presented previously \cite{WIN}. In this section we briefly recount the experimental technique, focusing on the experimental uncertainties.

\subsection{\label{sec:ExpTech} Experimental Technique}

A beam of $^8$B ions was implanted near the midplane of a 91~$\mu$m thick planar Si detector. An implanted source eliminates the possibility of energy loss outside the sensitive region of the detector, a systematic effect in all previous experiments. The detector thickness was just sufficient to stop $\alpha$ particles emitted with the highest possible energy (about 8.5~MeV). Thus the full energy of both $\alpha$ particles was detected while the positrons, usually close to minimum ionizing, deposited only a small amount of energy. The systematic effect of positron energy was further reduced with a coincidence detector, selecting events where the positron trajectories were close to normal to the Si detector surface. The system was calibrated using a beam of $^{20}$Na which was also implanted near the detector midplane. The $^{20}$Na decays with 20\% probability to $\alpha$ unstable levels in $^{20}$Ne, providing calibration lines of well-known energy \cite{ISO}. An external $^{228}$Th $\alpha$ source was used to provide additional calibration lines.

The experiment used the ATLAS superconducting linear accelerator at the Argonne National Laboratory. The $^8$B (t$_{1/2}$=770$\pm$3~ms) beam was produced by the In-Flight Technique \cite{HAR} using the $^3$He($^6$Li,$^8$B)n reaction. The primary $^6$Li beam, with energy 36.4 MeV, bombarded a 3.5 cm long gas cell filled with 700~mbar $^3$He and cooled to 82 K. The gas cell was separated from the beam-line vacuum by titanium windows. The pressure and temperature in the cell were held constant to $\pm$1\%.

\begin{figure}
\includegraphics[width=0.35\textwidth]{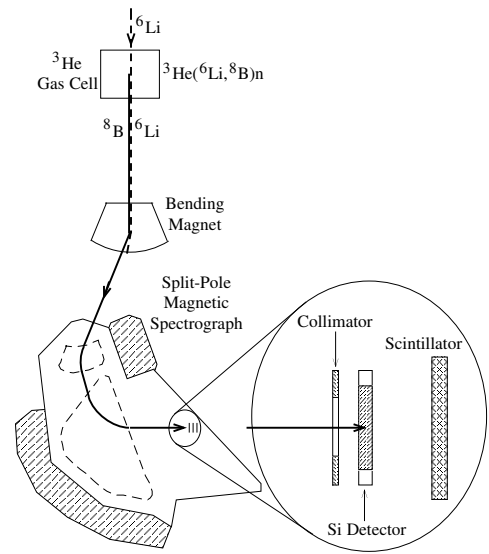}
\caption{\label{fig:apparatus} The experimental setup used to produce the $^8$B ($^{20}$Na) beam, separate it from the primary $^6$Li ($^{19}$F) beam, and select ions with energy 27.3~MeV (170~MeV) for implantation into the Si detector (not to scale).}
\end{figure}

Fully stripped $^8$B products were separated from the primary beam with a 22$^\circ$ bending magnet, and transported through an Enge Split Pole spectrograph. A gas-filled detector located in the focal plane of the spectrograph \cite{FPD} identified the $^8$B products by mass, nuclear charge, and energy. The spectrograph was then adjusted so that $^8$B ions with energies of 27.3$\pm$0.2~MeV were incident on the planar Si detector with a 150~mm$^2$ active area (13.8~mm diameter), located adjacent to the focal plane detector. An 11~mm diameter Ta collimator masked the edges of the detector. The beam was cycled (1.5 sec on/1.5 sec off) and data taken only during the beam-off cycles. The average implantation rate was 3~$^8$B ions/sec, and 4.5$\times$10$^{5}$ decays were observed over six days.

Using the gas-filled focal plane detector, the $^8$B beam purity was measured to be about 10$^{-3}$. A portion of the low-energy tail of the primary $^6$Li beam, as well as $\alpha$ particles, deuterons, and protons with the proper magnetic rigidity to traverse the spectrograph, accounted for the remainder of ions incident on the detector. A $^6$Li beam incident on $^3$He cannot produce any $\beta$ delayed particle emitters other than $^8$B, which could create a background during the beam-off data acquisition cycles.  No products resulting from possible interactions between $^6$Li and the titanium windows of the gas-cell were observed with the proper rigidity to be incident on the Si detector.

The $\beta$ particle detector, located 12~mm behind the Si detector, was a 25~mm diameter $\times$ 2~mm thick plastic scintillator coupled by a lightguide to a Hamamatsu R647 photomultiplier tube.  The detector identified a subset of events where the positron from the $^8$B decay exited the Si detector with a trajectory within 30$^\circ$ to normal. Roughly 16\% of the observed events occurred in coincidence with a count in the $\beta$ detector, consistent with expectations from detector geometry. The Si/scintillator detector system was cooled to -5$^{\circ}$C. A schematic representation of the apparatus is shown in Fig. \ref{fig:apparatus}.

\begin{figure}
\includegraphics[width=0.5\textwidth]{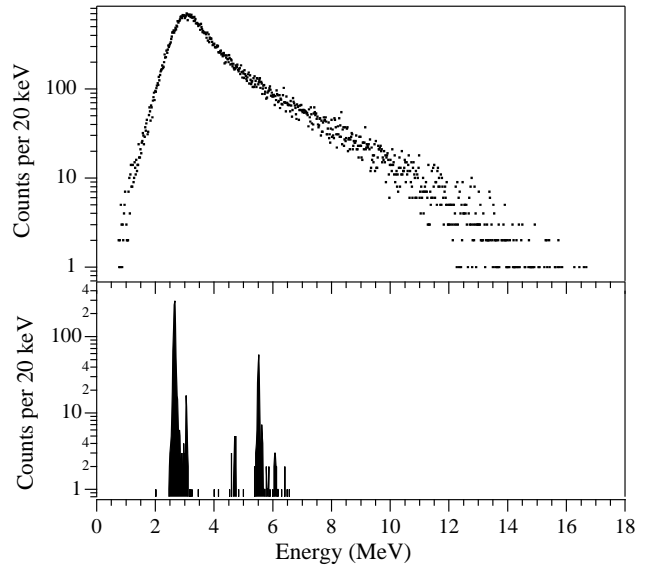}
\caption{\label{fig:alpha} The measured $^8$B $\beta^+$ delayed two alpha spectrum shown with the $^{20}$Na $\beta^+$ delayed alpha lines used for calibration. The data shown here correspond to events coincident with the beta detector.}
\end{figure}

The calibration using implanted $^{20}$Na was performed immediately before the $^8$B run. The $^{20}$Na $\beta^+$ delayed alpha particles provided three calibration lines near the region of the $^8$B $\alpha$ spectrum peak, with energy releases of 2691.9$\pm$1.2, 3099.0$\pm$2.2, and 5544.0$\pm$2.8~keV \cite{ISO}. The $^{20}$Na (t$_{1/2}$=448$\pm$3~ms) beam was produced by using the $^{19}$F($^3$He,2n)$^{20}$Na reaction and separating fully stripped $^{20}$Na ions with energies of 170.0$\pm$1.5~MeV. A mylar degrader foil of thickness 85$\pm$4~$\mu$m in front of the detector slowed the ions prior to implantation. As in the $^8$B runs, the beam was cycled (1.0 sec on/1.0 sec off). An average implantation rate of 7 $^{20}$Na ions/min was achieved, and over one day 1.0$\times$10$^4$ decays were observed. The raw energy spectra from the $^8$B and $^{20}$Na runs are displayed in Fig. \ref{fig:alpha}.

The integrated incident flux on the Si detector, monitored by the spectrograph focal plane detector, was an order of magnitude below threshold for detector damage \cite{ORTEC}, and no gain variation from damage was expected. The gain was monitored with the centroid of the $^8$B $\alpha$ spectrum and was found to fluctuate within $\pm$0.25\%, corresponding to $\pm$7~keV at the spectrum peak. The fluctuations are about two times larger than the statistics. External $\alpha$ particle sources were not reliable for monitoring gain shifts because of the accretion of residual gas onto the cooled Si detector, degrading the $\alpha$ particle energies by 10-20~keV over the course of the seven day run. The accreted gas was not sufficient to appreciably degrade the $^8$B and $^{20}$Na beams incident on the detector, and did not affect implantation depth.

The electrostatic sweeper used to stop the beam was not perfectly efficient, allowing a weak beam during the counting cycles.  Protons with energies near 8.7~MeV, produced in reactions from the primary beams, had the right rigidity to traverse the spectrograph and hit the Si detector. The 8.7~MeV protons passed through the Si detector and into the $\beta$ detector, producing a peak near 800~keV in the coincidence data. These protons were rejected based on the large pulses observed in the $\beta$ detector, which were much larger than the pulses from minimum ionizing positrons. Any ions heavier than protons with the proper rigidity to reach the Si detector were stopped \cite{TRIM} and rejected by the coincidence requirement.

The analysis provided here is slightly improved over Ref. \cite{WIN}. Proton events were removed by cuts, as noted above. Random coincidences were removed with cuts on the time spectrum recorded between the $\beta$ detector start and Si detector stop signal. Additional cuts on the time spectrum eliminated a small amount of background from external $\beta$ decay activity.

\subsection{\label{sec:ExpUnc} Experimental Uncertainties}

The largest sources of experimental uncertainties in the $\alpha$ spectrum measurement, affecting the $^8$B neutrino spectrum, are:  (1) The temporal gain variation observed over the seven days of data collection. (2) The uncertainty in correcting the energy deposited by positrons, which includes the uncertainty in implantation depth of the $^8$B and $^{20}$Na ions. (3) The uncertainty in the energy scale calibration. 

As noted, the nonstatistical gain variation over the seven day run was of magnitude $\pm$0.25\% which corresponds to a $\pm$7~keV uncertainty at the peak of the $\alpha$ spectrum. A gain correction was not applied, instead an uncertainty is included in the energy scale. This is the dominant source of uncertainty in the measurement.

The depth and distribution of the implanted ions was estimated using the TRIM Monte Carlo simulation \cite{TRIM}. The uncertainty is taken as $\pm$4.6\%, the average deviation of TRIM estimates from measured stopping power \cite{TRIM}. For $^8$B ions of incident energy 27.3$\pm$0.2~MeV, TRIM predicts an average implantation depth is 42.2$\pm$2.0~$\mu$m. The full-width half maximum of the implantation depth is 0.7 $\mu$m. For $^{20}$Na ions of energy 170.0$\pm$1.5~MeV, first passing through the mylar degrading foil of thickness 85$\pm$4~$\mu$m, TRIM predicts an average implantation depth of 48$\pm$6~$\mu$m for $^{20}$Na ions, with a full width half maximum of 1.3~$\mu$m. 

Uncertainties in ion implantation depth correspond to uncertainties in energy deposited by positrons. On average, minimum ionizing positrons deposit 0.6 keV/$\mu$m in Si, so that in the case of $^8$B ($^{20}$Na) the uncertainty in implantation depth corresponds to an energy uncertainty of $\pm$1.2 keV ($\pm$3.6 keV).

The positron energy loss in the Si detector was estimated using the EGSnrc simulation \cite{EGS}. Simulations account for detector geometry and positron energy spectra and assume the ranges of ion implantation depths discussed previously. Probability distributions for energy loss by positrons were obtained for the subset of data associated with a coincidence count in the $\beta$ detector, and for the total data set. The uncertainty associated with these simulations was estimated by comparing the total $^8$B data set to the coincidence data set. The effect of the positron correction lowered the $\alpha$-spectrum peak of the total data set by 55 keV, and the peak of the coincidence data set by 24 keV. After the correction, the peaks of the two data sets agreed to within 2 keV. The uncertainty associated with the simulation is thus assigned as $\pm$2~keV. The use of the total data set to estimate uncertainty in positron energy loss was not compromised by the beam leakage background, since beam particles with the proper rigidity to hit the detector had energies far from the $^8$B spectrum peak.

The average pulse height defect of the recoil $^{16}$O nuclei, which carry one fifth of the energy of the $\alpha$ disintegrations following $^{20}$Na decay, has been directly measured for $^{16}$O nuclei in the energy range of interest \cite{LGW}. The correction is 40-50~keV for the various $^{20}$Na alpha lines, with an uncertainty of $\pm$5~keV. The TRIM Monte Carlo simulation \cite{TRIM} was used to model the ionization energy loss of $^{16}$O in silicon, and agreed within 2~keV with the average values of ionization loss observed in \cite{LGW}. We have applied the TRIM results, scaled by 2~keV to agree with the experimental results, to approximate the pulse height spectrum of $^{16}$O nuclei in a silicon detector.

\begin{figure}
\includegraphics[width=0.5\textwidth]{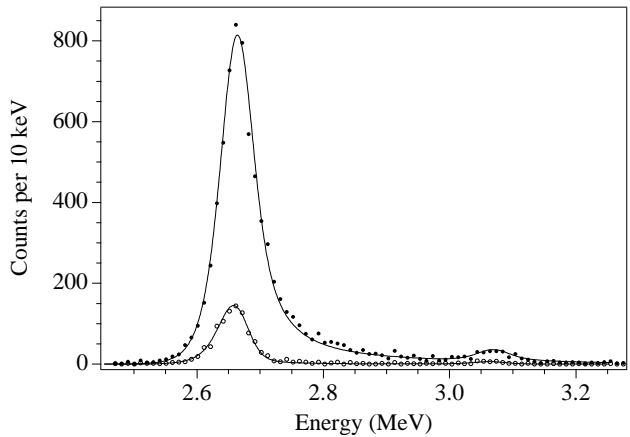}
\caption{\label{fig:sodium} 
Fits to the $^{20}$Na calibration lines. The open circles indicate the coincidence data set, while the solid circles indicate the total data set, i.e. no coincidence requirement. The curves show the best fit function, described in the text.}
\end{figure}  

The energy spectrum from the $^{20}$Na decay was used to calibrate the energy scale. The calibration lines were fit to the pulse height spectrum predicted by TRIM, convoluted with the positron energy loss distributions and a Gaussian component to approximate detector noise. The position and amplitude of the lines were free parameters, as well as the Gaussian width. Results of the fit to two of the lines, resulting from $^{20}$Na $\beta^+$ decays which led to $\alpha$ energy releases of 2691.9$\pm$1.2 and 3099.0$\pm$2.2~keV, are shown in Fig. \ref{fig:sodium} for both the total and coincidence data sets.

\begin{figure}
\includegraphics[width=0.5\textwidth]{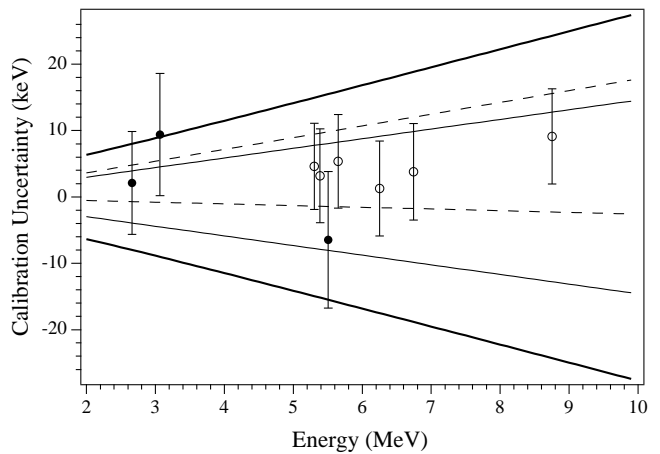}
\caption{\label{fig:calibration} Residuals from the calibration process. The solid circles show the residuals for the three $^{20}$Na calibration lines to the best linear fit (energy vs. ADC channel). The open circles correspond to the six external $^{228}$Th source alpha lines, which were not used in the calibration fit. The thin solid curves are the 1$\sigma$ error bands associated with the $^{20}$Na calibrations. The dashed curves are the 1$\sigma$ error bands of a separate calibration from the $^{228}$Th source. The thick solid curves show the total 1$\sigma$ uncertainty in the energy scale, which is significantly larger than the calibration uncertainty alone due to temporal gain variation.}
\end{figure}

Pulser tests performed before and during the data collection period indicate a negligible quadratic component in the relationship between pulse height and ADC bin, so the $^{20}$Na calibration was performed using a two parameter linear fit to the three dominant lines along with the zero energy ADC channel precisely determined by pulser tests. The external $^{228}$Th source emitted $\alpha$ particles at six distinct energies, 5.341, 5.423, 5.686, 6.288, 6.779, and 8.784~MeV \cite{ISO} and was used to perform an independent calibration. Data used for the $^{228}$Th calibration was taken immediately after the detector was placed in vacuum and cooled, before an appreciable amount of residual gas condensed on the detector surface. The $\alpha$ particle energies were corrected for energy loss in the source and the measured 27$\pm$4~$\mu$g/cm$^2$ detector dead layer. The magnitude of the corrections for the various lines was 31-38~keV, with a characteristic uncertainty of 4-5~keV. A comparison of the residuals from the two calibrations is shown in Fig. \ref{fig:calibration}. The figure also shows the total uncertainty in the energy scale, dominated by the temporal gain variation.

\section{\label{sec:AlphaFits}R-matrix Representation of the Alpha Spectrum}

The many-level R-matrix formalism has previously been used to parametrize data from nuclear processes involving $^8$Be, in particular the alpha spectrum following $^8$B $\beta^+$ decay \cite{BAR1,WAR,BAR2,ORT,BA}. The application of the R-matrix to $\beta$ decay is an approximation, and the physical significance of R-matrix fit parameters is not clear.

In principle, it is possible to deduce the $^8$B $\beta^+$ decay strength function directly from the measured $\alpha$ spectrum without resorting to R-matrix formalism. The R-matrix approach, however, gives a good fit to the observed $\alpha$ spectrum and provides a convenient method for propagating systematic uncertainties in the $\alpha$ spectrum to the neutrino spectrum. Systematic uncertainties in the $\alpha$ spectrum dominate the statistical uncertainties, justifying the representation of the data by a smooth function.

\subsection{\label{sec:RLevels}Energy Levels in $^8$Be}

In the R-matrix approximation, the $\beta^+$ decay of $^8$B proceeds with varying strength through a region of interfering nuclear states in $^8$Be which immediately decay into $\alpha$ particles. Each state is characterized by an excitation energy, $\mathcal{E}_j$, a $\textit{reduced width}$, $\gamma_j$, and a $\beta^+$ decay strength quantified by Fermi and Gamow-Teller matrix elements, $\mathcal{M}_{Fj}$ and $\mathcal{M}_{GTj}$.

As discussed in Sec. \ref{sec:Intro}, only 2$^+$ states in $^8$Be are considered. A numerically accurate R-matrix fit to the observed $\alpha$ spectrum requires the three 2$^+$ states in $^8$Be shown in Fig. \ref{fig:levels}, as well as one phenomenological $\textit{background state}$ approximating the combined effect of all higher-lying 2$^+$ states. It has been shown \cite{WAR,BA} that R-matrix fits using only these four states were sufficient to describe previous experimental data. We have repeated the analysis discussed in Ref. \cite{WAR}, which explicitly included a greater number of 2$^+$ states, and have verified that the four state R-matrix approximation is sufficient to describe the $\alpha$ spectrum reported here.

The state labeled (j=1), with excitation energy near 3~MeV and width of about 1.5~MeV, is responsible for the peak of the observed $\alpha$ spectrum. The excitation energy, $\mathcal{E}_1$, and reduced width, $\gamma_1$, are considered free fit parameters. Shell-model considerations \cite{BARsh} indicate no significant Fermi decay strength to this level, as discussed in Ref. \cite{MGG} which reports measurements of the $\beta$-$\nu$-$\alpha$ correlations in $^8$B and $^8$Li consistent with a pure Gamow-Teller decay. We take the Fermi decay strength to vanish, $\mathcal{M}_{F1}$=0, while the Gamow-Teller matrix element, $\mathcal{M}_{GT1}$, is a free parameter.

The next two states (j=2,3) form a nearly degenerate doublet with excitation energies 16.626(3) and 16.922(3)~MeV \cite{ISO} which are well known to be almost maximally mixed in isospin. We describe the isospin mixing of the doublet using the standard formulation \cite{BAR1} and consider the energy eigenstates $\psi_2$ and $\psi_3$ in terms of the isospin eigenstates $\phi_{A}$~(T=0) and $\phi_{B}$~(T=1),
\begin{equation}
\label{eq:isomix}
\psi_{2}=\alpha\phi_{A}+\beta\phi_{B},\ \ \  \psi_{3}=\beta\phi_{A}-\alpha\phi_{B},
\end{equation}
where $\alpha$ and $\beta$ are mixing parameters with $\alpha^2+\beta^2=1$. Since $\alpha$ decays from a T=1 state are forbidden, the parameters $\alpha$ and $\beta$ may be approximated from the level widths,
\begin{equation}
\alpha^{2}=\Gamma_{2}/(\Gamma_{2}+\Gamma_{3}), \ \ \ 
\beta^{2}=\Gamma_{3}/(\Gamma_{2}+\Gamma_{3}).
\end{equation}

An accurate R-matrix description of the alpha spectrum requires $\alpha$,$\beta>$0. The energies, $\mathcal{E}_2$ and $\mathcal{E}_3$, and reduced widths, $\gamma_2$ and $\gamma_3$, of the doublet are well constrained by $\alpha$-$\alpha$ scattering experiments \cite{AA} and are held constant.

The decomposition of the doublet into its component isospin eigenstates allows a simplified description of the Fermi and Gamow-Teller strengths. The T=0 state, $\phi_{A}$, has a Gamow-Teller strength treated as a free parameter, $\mathcal{M}_{GTA}$. The T=1 state, $\phi_{B}$, is the isospin analog of the $^8$B and $^8$Li ground states and is populated by Fermi decay with a strength given by the superallowed Fermi matrix element, $\mathcal{M}_{FB}$=$\sqrt{2}$. The Gamow-Teller decay to the T=1 component, or Fermi decay to the T=0 component, may be nonzero due to isospin breaking but has been estimated to be negligible \cite{KUR} in this context, as discussed in Ref. \cite{WAR}. Hence we take $\mathcal{M}_{GTB}$=0 and $\mathcal{M}_{FA}$=0. The matrix elements of the isospin eigenstates are then related to the matrix elements of the energy eigenstates by Eq. \ref{eq:isomix}, 
\begin{equation}
\mathcal{M}_{F2}=\beta \mathcal{M}_{FB} ,\ \ \  \mathcal{M}_{F3}=-\alpha \mathcal{M}_{FB},
\end{equation}
and
\begin{equation}
\mathcal{M}_{GT2}=\alpha \mathcal{M}_{GTA} , \ \ \ \mathcal{M}_{GT3}=\beta \mathcal{M}_{GTA}.
\end{equation}

The background state labeled (j=4) has an excitation energy held fixed to a value near that used in recent works \cite{ORT,BA,WAR}, $\mathcal{E}_4$=37.0. The parameter $\mathcal{E}_4$ could be allowed to float, but the quality of the fit is very weakly dependent on its value. The reduced width, $\gamma_4$, and the Gamow-Teller matrix element, $\mathcal{M}_{GT4}$, are free parameters. The Fermi strength is taken to be negligible, $\mathcal{M}_{F4}$=0. 

\subsection{\label{sec:RForm}Form of the R-matrix Function}

The R-matrix approach gives a parametrization of the $^8$B $\beta^+$ decay strength function, indicating the probability that $^8$Be is populated at a given excitation energy, E$_x$. The function takes the form \cite{BAR1}

\begin{equation}
\label{eq:RMatrixEq}
\frac{dN}{dE_x} = \left(\frac{Nt_{1/2}}{6166} \right)f_{\beta}(E_x)
\Big(a^2(E_x)+c^2(E_x)\Big).
\end{equation}

Here N is the total number of observed decays,
\begin{equation}
N=\int \frac{dN}{dE_x} dE_x.
\end{equation}

The lifetime of $^8$B, t$_{1/2}$, is 770$\pm$3~msec \cite{ISO}.  The unitless integrated phase space available to the $\beta$ decay leptons, $f_\beta$(E$_x$), including the Fermi function and “outer” radiative corrections, has been evaluated according to the parametrization given by Wilkinson and Macefield \cite{WiMa}. The Fermi and Gamow-Teller matrix elements, a(E$_x$) and c(E$_x$), are parametrized by 

\begin{widetext}
\begin{equation}
a^2(E_x) = \frac{P(E_x)}{\pi} \left(\frac{\left|\sum_{j=1}^{4}\frac{\mathcal{M}_{Fj}\gamma_{j}}{\mathcal{E}_{j}-E_x}\right|^2}{\left|1-\Big(S(E_x)-B+iP(E_x)\Big)\sum_{j=1}^{4}\frac{\gamma_j^{2}}{\mathcal{E}_{j} -E_x}\right|^{2}}\right)
\end{equation}
and
\begin{equation}
\label{eq:c}
c^2(E_x) = \frac{P(E_x)}{\pi} \left(\frac{\left|\sum_{j=1}^{4}\frac{\mathcal{M}_{GTj}\gamma_{j}}{\mathcal{E}_{j}-E_x}\right|^2}{\left|1-\Big(S(E_x)-B+iP(E_x)\Big)\sum_{j=1}^{4}\frac{\gamma_j^{2}}{\mathcal{E}_{j} -E_x}\right|^{2}}\right).
\end{equation}
\end{widetext}

The P(E$_x$) and S(E$_x$) are the penetrability and shift factor arising from the regular and irregular solutions of the Coulomb equation of L=2 $\alpha$ particles, defined in Ref. \cite{LT}. As in previous works \cite{WAR,ORT,BA}, we evaluate the Coulomb functions at matching radius r$_c$=4.5~fm, and choose the boundary condition, B, to be S($\mathcal{E}_1$).

\subsection{\label{sec:RResults}Application to Data Set}

\begin{figure}
\includegraphics[width=0.5\textwidth]{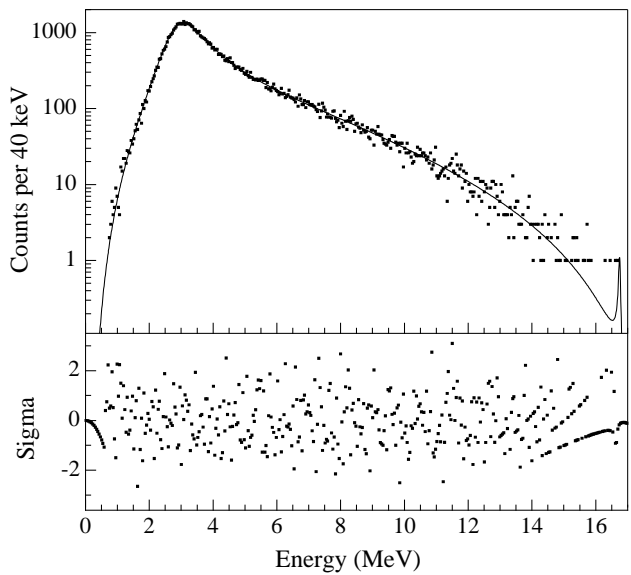}
\caption{\label{fig:rfit} Top panel: R-matrix fit to the observed decay spectrum. Bottom panel: Residuals to the fit, scaled by the square root of the fit value.}
\end{figure}

In cases where the $^8$B decays at rest, the recoil of the daughter $^8$Be nucleus will cause the total energy spectrum of the emitted $\alpha$ particles to deviate from the $\beta^+$ decay strength function given in Eq. \ref{eq:RMatrixEq}. For a given excitation energy of $^8$Be, the recoil energy distribution is exactly calculable and takes an average value of 7~keV at the most probable excitation energy near 3.0~MeV.

In addition to accounting for the $^8$Be recoil, the strength function (Eq. \ref{eq:RMatrixEq}) must be convoluted with the probability distribution of energies deposited by the positron, discussed in Sec. \ref{sec:Alpha}.
The detector line shape, approximated as a Gaussian with width 25~keV, determined by fits to the $^{20}$Na data sets, was also included but had a negligible impact on the fit due to the large width of the $\alpha$ spectrum.

\begin{figure}
\includegraphics[width=0.5\textwidth]{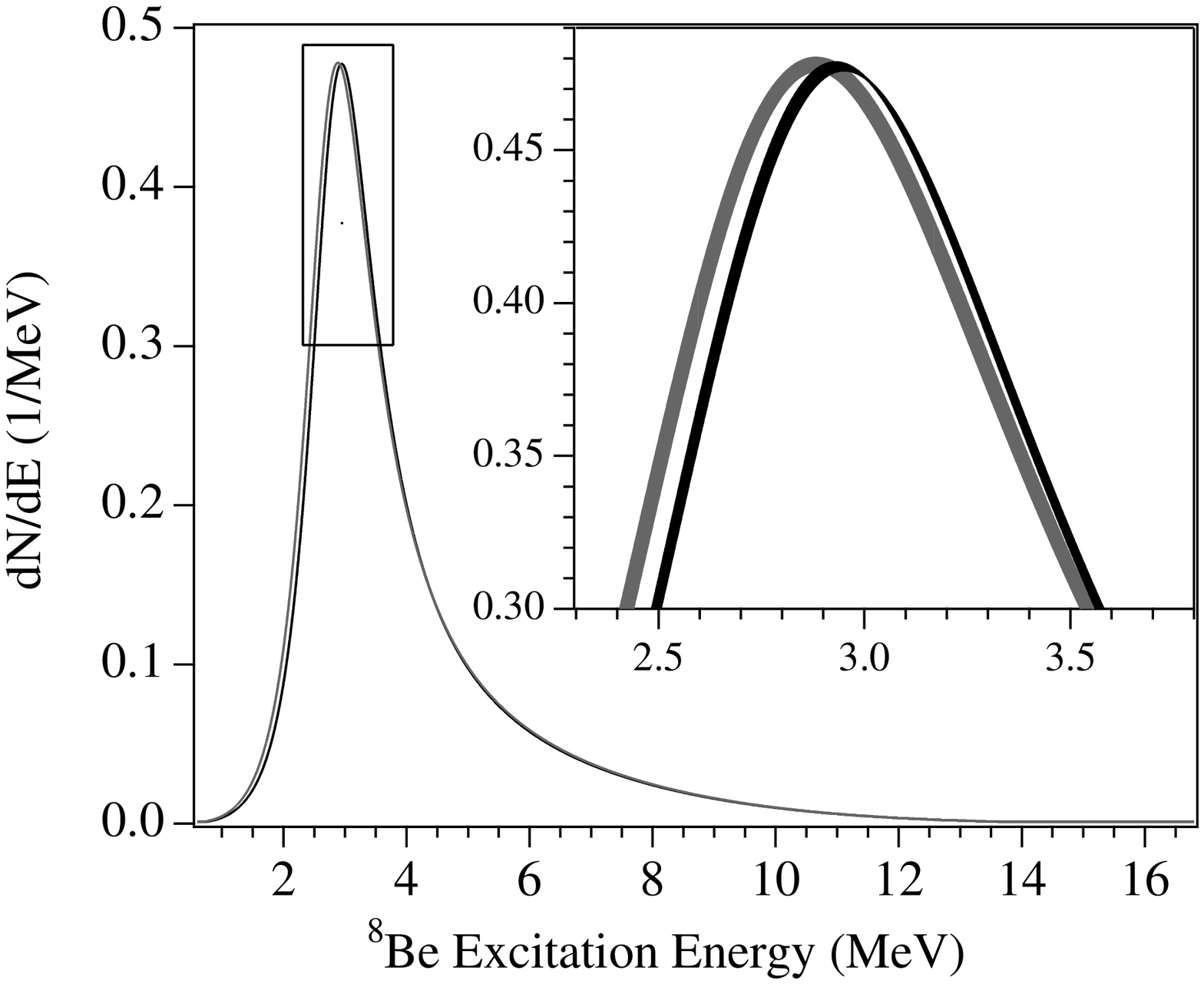}
\caption{\label{fig:compare}
The $^8$B $\beta^+$ decay strength function, determined by R-matrix fits to the $^8$B alpha spectrum presented in this work (black) and in Ortiz $\textit{et al.}$ \cite {ORT} (grey). The spectra are scaled to share the same peak height. The insert shows the locations of the spectrum peaks, on which the neutrino spectrum is highly dependent. The width of the lines in the insert indicate the magnitude of the $\pm$1$\sigma$ experimental uncertainties. The thin feature in the blue curve arises because the dominant uncertainty is a multiplicative factor in the energy scale.
}
\end{figure}

\begin{table}
\caption{\label{tab:Rlevels}Values of R-matrix parameters determined by a fit to the coincidence $\alpha$ spectrum data, using a matching radius of r$_c$=4.5~fm.}
\begin{ruledtabular}
\begin{tabular}{lr}
Parameter & Value\\
\hline
$\mathcal{E}_1$ & 3.043 MeV\\
$\mathcal{E}_2$ & 16.626 MeV\footnotemark[1]\\
$\mathcal{E}_3$ & 16.922 MeV\footnotemark[1]\\
$\mathcal{E}_4$ & 37.0 MeV\footnotemark[1]\\
$\gamma_1^2$ & 1.087 MeV\\
$\gamma_2^2$ & 10.96 keV\footnotemark[1]\\
$\gamma_3^2$ & 7.42 keV\footnotemark[1]\\
$\gamma_4^2$ & 5.619 MeV\\
$\mathcal{M}_{GT1}$ & -0.1462\\
$\mathcal{M}_{GTA}$ & 2.423\\
$\mathcal{M}_{GT4}$ & -0.1320\\
$\mathcal{M}_{FB}$ & $\sqrt{2}$\footnotemark[1]\\
$\mathcal{M}_{GTB}$, $\mathcal{M}_{F1}$, $\mathcal{M}_{FA}$, $\mathcal{M}_{F4}$ & 0\footnotemark[1]\\
\end{tabular}
\end{ruledtabular}
\footnotetext[1]{These parameters were held constant during the fitting procedure.}
\end{table}

The $\alpha$ spectrum data was fit using the log-likelihood minimization function \cite{BC}. The best fit gave $\chi^2$/dof=3249.7/3376, indicating a satisfactory fit. The best fit parameters are given in Table \ref{tab:Rlevels}, and the best fit is compared to the data in Fig. \ref{fig:rfit}. The strength function is presented in numerical form in Table \ref{tab:strengthtable}.

\begin{table*}[!]
\caption{\label{tab:strengthtable}The $^8$B $\beta^+$ decay strength function, as determined by fitting the experimental $\alpha$ spectrum to Eq. \ref{eq:RMatrixEq}. The strength function is normalized to 1000 when integrated with respect to MeV. Note that the energy spacing of data points varies to allow a more detailed description of the strength function near the peak. Uncertainties in the function are well approximated by deforming the energy scale. The 1$\sigma$ uncertainty in the energy scale is characterized by a multiplicative factor of 1$\pm$(0.275\%), corresponding to about 8~keV at the spectrum peak, added in quadrature with a constant offset of 3~keV.}
\begin{ruledtabular}
\begin{tabular}{cccccccccccccccc}
E$_x$ & dN/dE$_x$ & E$_x$ & dN/dE$_x$ & E$_x$ & dN/dE$_x$ & E$_x$ & dN/dE$_x$ & E$_x$ & dN/dE$_x$ & E$_x$ & dN/dE$_x$ & E$_x$ & dN/dE$_x$ & E$_x$ & dN/dE$_x$\\
\hline
0.00 &  0.000 & 1.90 & 67.991 & 2.60 & 362.922 & 3.06 & 462.243 & 3.52 & 315.256 & 4.70 & 117.714 & 8.00 & 24.113 & 12.60 &  2.349 \\ 
 0.10 &  0.000 & 1.95 & 78.048 & 2.62 & 374.309 & 3.08 & 457.936 & 3.54 & 309.037 & 4.80 & 110.362 & 8.20 & 22.147 & 12.80 &  2.054 \\ 
 0.20 &  0.004 & 2.00 & 89.464 & 2.64 & 385.390 & 3.10 & 453.156 & 3.56 & 302.941 & 4.90 & 103.693 & 8.40 & 20.333 & 13.00 &  1.788 \\ 
 0.30 &  0.024 & 2.05 & 102.397 & 2.66 & 396.095 & 3.12 & 447.953 & 3.58 & 296.971 & 5.00 & 97.622 & 8.60 & 18.655 & 13.20 &  1.548 \\ 
 0.40 &  0.081 & 2.10 & 117.004 & 2.68 & 406.353 & 3.14 & 442.380 & 3.60 & 291.129 & 5.10 & 92.074 & 8.80 & 17.103 & 13.40 &  1.333 \\ 
 0.50 &  0.207 & 2.15 & 133.440 & 2.70 & 416.096 & 3.16 & 436.485 & 3.62 & 285.417 & 5.20 & 86.986 & 9.00 & 15.667 & 13.60 &  1.141 \\ 
 0.60 &  0.440 & 2.20 & 151.844 & 2.72 & 425.257 & 3.18 & 430.314 & 3.64 & 279.834 & 5.30 & 82.305 & 9.20 & 14.334 & 13.80 &  0.970 \\ 
 0.70 &  0.832 & 2.25 & 172.320 & 2.74 & 433.776 & 3.20 & 423.911 & 3.66 & 274.380 & 5.40 & 77.984 & 9.40 & 13.100 & 14.00 &  0.818 \\ 
 0.80 &  1.443 & 2.30 & 194.916 & 2.76 & 441.597 & 3.22 & 417.319 & 3.68 & 269.056 & 5.50 & 73.984 & 9.60 & 11.956 & 14.20 &  0.685 \\ 
 0.90 &  2.348 & 2.32 & 204.543 & 2.78 & 448.671 & 3.24 & 410.576 & 3.70 & 263.859 & 5.60 & 70.270 & 9.80 & 10.896 & 14.40 &  0.568 \\ 
 1.00 &  3.640 & 2.34 & 214.498 & 2.80 & 454.958 & 3.26 & 403.717 & 3.72 & 258.789 & 5.70 & 66.814 & 10.00 &  9.915 & 14.60 &  0.466 \\ 
 1.10 &  5.435 & 2.36 & 224.770 & 2.82 & 460.423 & 3.28 & 396.777 & 3.74 & 253.845 & 5.80 & 63.588 & 10.20 &  9.006 & 14.80 &  0.378 \\ 
 1.20 &  7.878 & 2.38 & 235.344 & 2.84 & 465.045 & 3.30 & 389.783 & 3.76 & 249.024 & 5.90 & 60.572 & 10.40 &  8.166 & 15.00 &  0.302 \\ 
 1.30 & 11.155 & 2.40 & 246.203 & 2.86 & 468.808 & 3.32 & 382.763 & 3.78 & 244.323 & 6.00 & 57.744 & 10.60 &  7.389 & 15.20 &  0.238 \\ 
 1.40 & 15.507 & 2.42 & 257.323 & 2.88 & 471.708 & 3.34 & 375.742 & 3.80 & 239.742 & 6.20 & 52.589 & 10.80 &  6.672 & 15.40 &  0.185 \\ 
 1.50 & 21.242 & 2.44 & 268.676 & 2.90 & 473.750 & 3.36 & 368.740 & 3.90 & 218.531 & 6.40 & 48.008 & 11.00 &  6.011 & 15.60 &  0.140 \\ 
 1.55 & 24.750 & 2.46 & 280.230 & 2.92 & 474.946 & 3.38 & 361.778 & 4.00 & 199.905 & 6.60 & 43.909 & 11.20 &  5.402 & 15.80 &  0.104 \\ 
 1.60 & 28.761 & 2.48 & 291.947 & 2.94 & 475.318 & 3.40 & 354.870 & 4.10 & 183.532 & 6.80 & 40.221 & 11.40 &  4.842 & 16.00 &  0.076 \\ 
 1.65 & 33.347 & 2.50 & 303.785 & 2.96 & 474.895 & 3.42 & 348.031 & 4.20 & 169.102 & 7.00 & 36.886 & 11.60 &  4.328 & 16.20 &  0.056 \\ 
 1.70 & 38.584 & 2.52 & 315.694 & 2.98 & 473.712 & 3.44 & 341.275 & 4.30 & 156.340 & 7.20 & 33.856 & 11.80 &  3.857 & 16.40 &  0.044 \\ 
 1.75 & 44.561 & 2.54 & 327.623 & 3.00 & 471.809 & 3.46 & 334.611 & 4.40 & 145.009 & 7.40 & 31.093 & 12.00 &  3.426 & 16.60 &  0.097 \\ 
 1.80 & 51.378 & 2.56 & 339.512 & 3.02 & 469.230 & 3.48 & 328.049 & 4.50 & 134.904 & 7.60 & 28.566 & 12.20 &  3.033 & 16.80 &  0.001 \\ 
 1.85 & 59.146 & 2.58 & 351.300 & 3.04 & 466.025 & 3.50 & 321.595 & 4.60 & 125.854 & 7.80 & 26.246 & 12.40 &  2.674 & 17.00 &  0.000 \\ 
 
\end{tabular}
\end{ruledtabular}
\end{table*}

The strength function reported here disagrees with the result of Ortiz $\textit{et al.}$ \cite{ORT}. For both measurements, the uncertainty in the inferred neutrino spectrum is dominated by systematic effects. Smooth R-matrix fits to the alpha spectra thus provide a convenient way to compare the two results. Fig. \ref{fig:compare} shows a comparison of the present results and a fit to the data of Ortiz $\textit{et al.}$ \cite{ORT}. Uncertainties in the Ortiz $\textit{et al.}$ curve are taken directly from Ref. \cite{ORT}.

\subsection{\label{sec:SysProp}Propagation of Systematic Uncertainties}

The R-matrix approach was used to propagate the systematic uncertainties in the $\alpha$ spectrum measurement to the neutrino spectrum. As discussed in Sec. \ref{sec:Alpha}, the 1$\sigma$ uncertainty in the energy scale is characterized by a multiplicative factor of 1$\pm$(0.275\%), corresponding to about 8~keV at the spectrum peak, added in quadrature with a constant offset of 3~keV. R-matrix fits were performed to the $\alpha$ spectrum using the $\pm$1$\sigma$ energy scales, and the resulting $\pm$1$\sigma$ strength functions were used to produce $\pm$1$\sigma$ neutrino spectra.

An additional uncertainty was imposed to account for the rapid drop off of the $\alpha$ spectrum at low energies, where statistics are not sufficient to determine the spectrum shape. The penetrability factor, P(E$_x$), is responsible for the drop off. The best R-matrix fit used P(E$_x$) calculated for a matching radius of 4.5~fm. We approximate the uncertainty at low energies by calculating P(E$_x$) at 4.0 and 5.0~fm, the $\pm$1$\sigma$ matching radii recommended in Ref. \cite{WAR}, and perform fits under these conditions. There is a strong dependence \cite{WAR} between the energy of the background state, $\mathcal{E}_4$ and matching radius, r$_c$, so the parameter $\mathcal{E}_4$ was allowed to float for these fits. We note that the variation of matching radius is a significant source of uncertainty only for neutrinos at very high (E$_\nu>$15~MeV) and low (E$_\nu<$0.5~MeV) energies.

\section{\label{sec:Recoil}Recoil Order Corrections to the Neutrino Spectrum}
\subsection{\label{sec:RecoilBack}Background}

A proper description of $^8$B $\beta^+$ decay includes recoil order effects which cause, for example, the energy spectra and angular correlations of decay particles to deviate from the allowed approximation. Deviations are of order E$_0$/m$_n$, where E$_0$ is the positron endpoint energy and m$_n$ is the nucleon mass. The $^8$B $\beta^+$ decay has a particularly large endpoint energy (most probable E$_0\approx$13~MeV) and a small Gamow-Teller strength (log ft = 5.6) for an allowed decay. Recoil order effects in $^8$B are thus large compared to other nuclear systems. 

Measurements of the radiative decay of the $^8$B isospin analog state in $^8$Be \cite{NAT,PSG,BG,DBW}, and of the angular correlation between $\beta$ and $\alpha$ particles emitted in the decays of $^8$B and $^8$Li \cite{TGL,TGC,MGG}, determine the recoil order matrix elements. These results were first explicitly applied to the neutrino spectrum in \cite{BH,NFC}, where they were found to contribute at the 5\% level. A more recent determination of the neutrino spectrum by Bahcall \textit{et al.} \cite{SNS} employed the same recoil order treatment as in Ref. \cite{BH}. Bahcall \textit{et al.} provided a conservative estimate of the uncertainty associated with the recoil order correction on the neutrino spectrum by setting the 3$\sigma$ uncertainty equal to size of the correction itself. A more recent determination of the neutrino spectrum by Ortiz \textit{et al.} \cite{ORT}, applied recoil order corrections very similar to those in Ref. \cite{BH}. 

The two most recent $\alpha$ spectrum measurements, by Ortiz \textit{et al.} \cite{ORT} and the one reported here, involved determinations of the energy scale significantly more precise than the measurements used by Bahcall \textit{et al.} \cite{SNS}. Also, a recent precision measurement of radiative decay in $^8$Be \cite{DBW} provides additional information on recoil order effects, but has not yet been applied to the $^8$B neutrino spectrum. In light of these recent experiments, recoil order effects are considered here with careful attention to the assignment of realistic uncertainties.

The Fermi matrix element plays a small role in the $\beta^+$ decay of $^8$B, contributing only to decays proceeding through the highest excitation energies in $^8$Be, as explicitly discussed in Ref. \cite{WAR}. These low energy $\beta^+$ decays have suppressed recoil order corrections, and produce neutrinos which have no impact on solar neutrino experiments. Consideration of the Fermi matrix element is thus omitted.

A model independent treatment of recoil order effects is given by Holstein \cite{HOL}, whose notation we adopt here. Matrix elements contributing to the $\beta$ decays of $^8$B and $^8$Li are denoted by $c$ (Gamow-Teller), $b$ (weak magnetism), $d$ (induced tensor), $f$, $g$ (vector second-forbidden), $j_2$, $j_3$ (axial second-forbidden), and $h$ (induced pseudoscalar). Since the decays proceed to the broad continuum in $^8$Be, the matrix elements should be considered as functions of the $^8$Be excitation energy, E$_x$. Previous determinations of the $^8$B neutrino spectrum \cite{BH,NFC,SNS,ORT} neglected this energy dependence.

\subsection{\label{sec:BetaSpec}Beta and Neutrino Energy Spectra}

The positron energy spectrum from an allowed decay proceeding between two energetically sharp nuclear states is given by
\begin{equation}
\frac{dN}{dE_\beta} \sim p_\beta E_\beta (E_0-E_\beta)^2 F(-Z,E_\beta) R(E_\beta,E_0) C(E_\beta,E_0).
\label{eq:betaspec}
\end{equation}
Here p$_\beta$ and E$_\beta$ are the momentum and total energy of the positron, and E$_0$ is the positron endpoint energy. F(-Z,E$_\beta$) is the Fermi function, which depends on the charge, Z, of the daughter nucleus and is negative for positron decays. The radiative corrections are contained in R(E$_\beta$,E$_0$), which will be discussed in Sec. \ref{sec:Neutrino}. The recoil order effects are contained in C(E$_\beta$,E$_0$), which has the form
\begin{widetext}
\begin{eqnarray}
\label{eq:betarecoil}
C(E_\beta,E_0) = 1 - \frac{2E_0}{3Am_n}\left(1+\frac{d}{c}-\frac{b}{c}\right) +\frac{2E_\beta}{3Am_n} \left(5-2\frac{b}{c}\right) -\frac{m_e^2}{3Am_nE_\beta} \left(2+\frac{d}{c}-2\frac{b}{c}-\frac{h}{c}\frac{E_0-E_\beta}{2Am_n}\right),
\end{eqnarray}
\end{widetext}
where A=8 is the mass number. In the case of $^8$B the recoil order matrix elements are  dependent on the $^8$Be excitation energy E$_x$ (E$_x$=$\Delta$-E$_0$), where $\Delta$=17.468~MeV is the total energy released in the $^8$B $\beta$-$\alpha$ decay chain. (This discussion of positron and neutrino energy spectra ignores, for the sake of simplicity, the kinetic recoil of the daughter nucleus. This effect is included in the numerical calculations.) The positron spectrum is calculated by integrating Eq. \ref{eq:betaspec} over all excitation energies in $^8$Be, weighted by the strength function determined in Sec. \ref{sec:AlphaFits}. The neutrino spectrum is obtained by the simple substitution E$_\nu$=E$_0$-E$_\beta$, and the application of different radiative corrections, discussed in Sec. \ref{sec:Neutrino}.

\subsection{\label{sec:RadWidth}Radiative Decay Measurements in $^8$Be}
The weak magnetism matrix element, $b$, exerts the greatest influence on the neutrino energy spectrum. Its value is best determined under the strong conserved vector current (CVC) hypothesis by measurements of the radiative decays of the $^8$B isospin analog state in $^8$Be which, as discussed in Sec. \ref{sec:AlphaFits}, is mixed between the two states of an energy doublet. The radiative decay is shown schematically in Fig. \ref{fig:levels}.

In previous experiments \cite{NAT,PSG,BG,DBW}, a $^4$He beam was directed at a $^4$He gas cell to excite the doublet in $^8$Be. The $^4$He($^4$He,$\gamma$)$^8$Be cross section was measured as a function of beam energy and angle of emission of the $\gamma$ ray. These measurements determine the widths of the isovector M1 and E2 transitions, $\Gamma^{T=1}_{M1}$ and $\delta_1$=$\Gamma^{T=1}_{E2}$/$\Gamma^{T=1}_{M1}$, as well as the widths of the isoscalar transitions, $\epsilon$=$\Gamma^{T=0}_{M1}$/$\Gamma^{T=1}_{M1}$ and $\delta_0$=$\Gamma^{T=0}_{E2}$/$\Gamma^{T=1}_{M1}$. The radiative widths are considered as functions of E$_x$. 

CVC relates the isovector radiative widths in $^8$Be to the vector recoil matrix elements contributing to $^8$B $\beta$ decay, $b$, $f$, and $g$,
\begin{equation}
b(E_x)=Am_n \sqrt{6 \Gamma^{T=1}_{M1}(E_x)/(\alpha E_\gamma^3)},
\end{equation}
\begin{equation}
f(E_x)=\frac{3}{10}\delta_1 b(E_x)
\end{equation}
\begin{equation}
g(E_x)=-\sqrt{\frac{2}{3}}\left(\frac{2Am_n}{E_0}\right)f(E_x).
\end{equation}
The isoscalar radiative widths are not related to $\beta$ decay from factors by CVC. 

\begin{table}
\caption{\label{tab:RadResults}Experimental determinations of the isovector and isoscalar M1 and E2 transition strengths. All quantities listed are integrated over final state excitation energies in $^8$Be.}
\begin{ruledtabular}
\begin{tabular}{lcr}
Observable&Experiment&Value\\
\hline
$\delta_1$ & Ref. \cite{NAT} (1975) & 0.045$\pm$0.027\\ 
 &  Ref. \cite{BG} (1978) & 0.14$\pm$0.03\footnotemark[1]\\ 
 &  Ref. \cite{DBW} (1995) & 0.01$\pm$0.03\\ 
\hline
$\delta_0$ & Ref. \cite{BG} (1978) & 0.26$\pm$0.03\footnotemark[1]\\ 
&  Ref. \cite{DBW} (1995) & 0.22$\pm$0.04\\ 
\hline
$\epsilon$ & Ref. \cite{BG} (1978) & 0.00$\pm$0.03\footnotemark[1]\\ 
 & Ref. \cite{DBW} (1995) & 0.04$\pm$0.02\\ 
\hline
$\Gamma^{T=1}_{M1}$ & Ref. \cite{PSG} (1977) & 4.1$\pm$0.6 eV\footnotemark[2]\\
 & Ref. \cite{BG} (1978)& 3.6$\pm$0.3 eV\footnotemark[2]\\
 & Ref. \cite{DBW} (1996) & 2.80$\pm$0.18 eV\\
\end{tabular}
\end{ruledtabular}
\footnotetext[1]{These values are based on a reanalysis of the original data, performed in Ref. \cite{DBW}. The original analysis contained an error in the kinematic treatment of the decay photon. See Ref. \cite{DBW} for details.}
\footnotetext[2]{The values for M1 width are based on a reanalysis of the original data, performed in Ref. \cite{DBW}, using the values of $\delta_1$ and $\delta_0$ obtained experimentally in Ref. \cite{DBW}.}
\end{table}

A summary of the experimental results is given in Table \ref{tab:RadResults}. The experimental results for the isoscalar contributions to the decay, $\epsilon$ and $\delta_0$, agree with each other and are of the same order as various shell model predictions compiled in Ref. \cite{DBW}.  The experimental values for $\delta_1$ from two of the experiments \cite{NAT,DBW} are in agreement, but differ from the results in \cite{BG} by about 3$\sigma$. The present work will use the more recent value of $\delta_1$ \cite{DBW} which indicates a negligible second-forbidden contribution to the decay, in agreement with shell model predictions. The early experimental determinations of $\Gamma_{M1}$ \cite {PSG,BG} disagree with the recent and most precise result \cite{DBW} by about 2$\sigma$. The recent result \cite{DBW} is in best agreement with $\beta$-$\alpha$ angular correlation experiments, as will be discussed later, and is adopted in this work.

\begin{figure}
\includegraphics[width=0.5\textwidth]{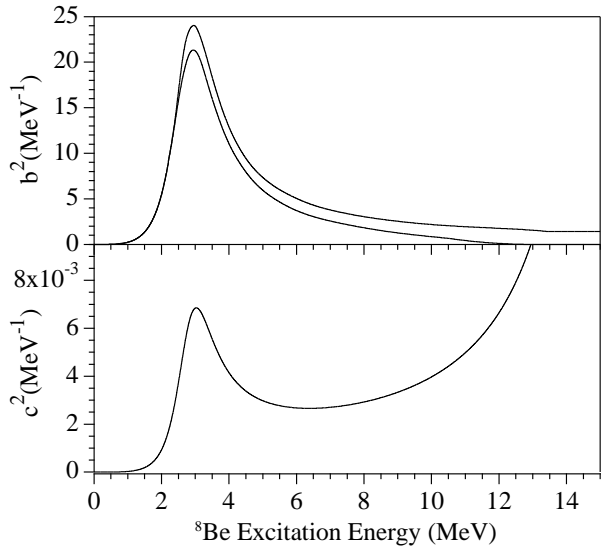}
\caption{\label{fig:matrix} The top panel shows the functional dependence of the weak magnetism matrix element, $b$(E$_x$), based on measurements of radiative decay in $^8$Be. The bands indicate 1$\sigma$ experimental uncertainties. The bottom panel shows the Gamow-Teller matrix element, $c$(E$_x$), based on fits to the alpha spectrum discussed in Sec. \ref{sec:AlphaFits}. The uncertainties in $c$(E$_x$) are comparable to the width of the line and are negligible in the context of recoil order corrections.}
\end{figure}

The matrix elements $b$(E$_x$) and $c$(E$_x$) have different functional dependences. This was first observed \cite{NAT,BG} through a comparison of the shapes of the final state distributions in $^8$Be following the $\alpha$ and $\gamma$ decays. The form of $b$(E$_x$) was later described \cite{DBW} using the R-matrix approach, which parametrized $b$(E$_x$) as an interfering sum of three different matrix elements, $\mathcal{M}_i$, to the three 2$^+$ levels in $^8$B shown in Fig. \ref{fig:levels}, 

\begin{widetext}
\begin{equation}
b^2(E_x) = \frac{P(E_x)}{\pi} \left(\frac{\left|\sum_{j=1}^{3}\frac{\mathcal{M}_{j}\gamma_{j}}{\mathcal{E}_{j}-E_x}\right|^2}{\left|1-\Big(S(E_x)-B+iP(E_x)\Big)\sum_{j=1}^{3}\frac{\gamma_j^{2}}{\mathcal{E}_{j} -E_x}\right|^{2}}\right)
\label{eq:b}
\end{equation}
\end{widetext}

The notations used here are identical to those in Sec. \ref{sec:AlphaFits}. We use the parameters reported in Ref. \cite{DBW} to determine $b$(E$_x$). The form of $c$(E$_x$) was given in Eq. \ref{eq:c}, and determined by fits to the $\alpha$ spectrum. We note that the R-matrix parameters appearing in both Eqs. \ref{eq:c} and \ref{eq:b} may take different values in the two expressions. The forms of $b$(E$_x$) and $c$(E$_x$) are shown in Fig. \ref{fig:matrix}. 

\subsection{\label{sec:BetaAlpha}$\beta$-$\alpha$ Angular Correlations}

The $\beta$-$\alpha$ angular correlations in the mirror decays of $^8$Li and $^8$B have been measured several times as a function of $\beta$ particle energy \cite{TGL,TGC,MGG}. Such measurements constrain the weak magnetism matrix element, $b$, as well as the induced tensor, $d$. The angular correlations take the form
\begin{equation}
N_{\mp}(\theta,E_{\beta},E_{x})=1+a_{\mp}(E_{\beta},E_{x})cos\theta+p_{\mp}(E_{\beta},E_{x})cos^2\theta,
\end{equation}
where the -(+) subscript refers to the $^8$Li($^8$B) decay, $\theta$ is the angle between the $\beta$ and $\alpha$ particles, and the factor v/c for the $\beta$ particle has been set equal to 1. The a$_{\mp}$ coefficients are dominated by kinematic considerations, while the p$_{\mp}$ coefficients are strongly dependent on recoil order contributions,
\begin{widetext}
\begin{equation}
p_{\mp}(E_{\beta},E_{x})=\frac{E_\beta}{2Am_nc}\left(\left(c-(d_I\mp d_{II})\pm b\right)\pm\frac{3}{\sqrt{14}}f\pm\sqrt{\frac{3}{28}}g\frac{\Delta-E_x-E_\beta}{Am_n}+\frac{3}{\sqrt{14}}j_2\frac{\Delta-E_x-2E_\beta}{2Am_n}-\frac{3}{\sqrt{35}}j_3\frac{E_\beta}{Am_n}\right),
\end{equation}
\end{widetext}
where $\Delta$=17.468~MeV is the total energy released in the $^8$B $\beta$-$\alpha$ decay chain.

Assuming isospin symmetry, taking the sum and difference of p$^-$ and p$^+$ produces cancellation between many of the mirror matrix elements of the $^8$B and $^8$Li decays. Corrections due to isospin breaking will be considered later. Defining $\delta_{\pm}$= p$_{-}\pm$p$_{+}$, dropping the vector matrix elements $f$ and $g$, integrating over excitation energy E$_x$ gives $\delta_{\pm}$ as a function of $\beta$ particle energy,

\begin{widetext}
\begin{equation}
\label{eq:deltaminus}
\delta_-(E_\beta)\frac{Am_n}{E_\beta} = \frac{\int b(E_x) c(E_x)(\Delta-E_x-E_\beta)^2dE_x}{\int c^2(E_x)(\Delta - E_x - E_\beta)^2dE_x}
\end{equation}
\begin{equation}
\label{eq:deltaplus}
\delta_+(E_\beta)\frac{Am_n}{E_\beta}=\frac{\int \left[c(E_x)-d(E_x)+\frac{3}{\sqrt{14}}j_2(E_x)\frac{\Delta-E_x-2E_\beta}{2Am_n}-\frac{3}{\sqrt{35}}j_3(E_x)\frac{E_\beta}{Am_n}\right]c(E_x)(\Delta-E_x-E_\beta)^2dE_x}{\int c^2(E_x)(\Delta - E_x - E_\beta)^2dE_x}
\end{equation}
\end{widetext}
where the second class contribution to the induced tensor has been omitted. This is consistent with existing data in the A=8 nuclear system \cite{DBW}, and with theoretical models which predict a second class current to contribute at a level below the current experimental sensitivity.

\begin{figure}
\includegraphics[width=0.5\textwidth]{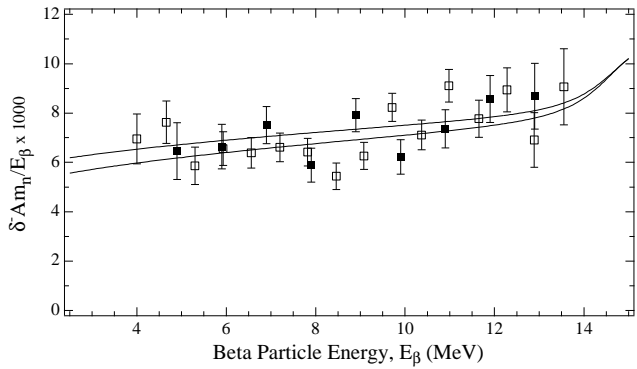}
\caption{\label{fig:deltaminus} The solid squares are experimental data on $\delta^-$ from $\beta$-$\alpha$ angular correlation measurements from Ref. \cite{TGC}, the open squares are from Ref. \cite{MGG}. The curves indicate the 1$\sigma$ error bands from the prediction for $\delta^+$ based on Eq. \ref{eq:deltaminus}, using the weak magnetism, $b$(E$_x$), and Gamow-Teller, $c$(E$_x$), matrix elements.}
\end{figure}

The matrix elements $b$(E$_x$) and $c$(E$_x$), determined previously, were applied to Eq. \ref{eq:deltaminus} to predict the $\delta^-$ observed in $\beta$-$\alpha$ angular correlation measurements \cite{TGL,TGC,MGG}. The predictions are compared to the experimental $\delta^-$ data graphically in Fig. \ref{fig:deltaminus}.
 
The level of agreement between the Eq. \ref{eq:deltaminus} prediction, based on the radiative decay and alpha spectrum data, and the $\beta$-$\alpha$ data sets was quantified by allowing the magnitude of $b$ to float by a multiplicative constant, $b\rightarrow \kappa b$, in Eq. \ref{eq:deltaminus}. The experimental data from the $\beta$-$\alpha$ correlation measurements was then used to determine the best fit value of $\kappa$. A value of $\kappa$ different than unity would indicate a disagreement between the radiative width data and the $\beta$-$\alpha$ angular correlation data. This approach was previously applied \cite{MGG,BG,NAT,DBW} with the motivation of testing CVC and searching for second-class currents. Here, the validity of CVC and the absence of second-class currents are assumed, and the test is performed to gauge the level of agreement between the two types of recoil order measurements.

The best fit to the $\delta^-$ angular correlation data from Ref. \cite{TGC} gave $\kappa$=1.06(4) with $\chi^2$/dof=7.7/8. The best fit to the data from Ref. \cite{MGG} gave $\kappa$=0.99(3) with $\chi^2$/dof=24.8/15, where the large $\chi^2$ value may be the result of the large point-to-point scatter of the data. The uncertainties in the data from Ref. \cite{MGG} were expanded by $\sqrt{\chi^2/dof}$ to account for this effect, and both data sets were fit simultaneously, yielding $\kappa$=1.014(26) with $\chi^2$/dof=24.6/24. The values of $\kappa$ obtained, consistent with unity, indicate agreement between the radiative width measurement \cite{DBW} and the $\beta$-$\alpha$ angular correlation measurements \cite{TGC,MGG}, and provide confidence in the extracted weak magnetism matrix element.

The experimental $\delta^+$(E$_\beta$) data is sensitive to the induced tensor matrix element, $d$. The effect of $d$ on the neutrino spectrum is much milder than that of $b$. The energy dependences of $b$(E$_x$) and $c$(E$_x$) were inferred directly from $\gamma$ and $\alpha$ spectrum measurements, respectively, but for $d$(E$_x$) there is no such experimental signal. The determination of the induced tensor is further complicated by the presence of the axial second-forbidden terms, $j_2$ and $j_3$, which appear in the expression for $\delta^+$, Eq. \ref{eq:deltaplus}. Fortunately, the influence of $d$ on the neutrino spectrum is sufficiently small that very conservative estimates of uncertainty may be imposed on $d$ without significantly inflating the total uncertainty of recoil order corrections.

The $\beta$ particle asymmetry from a polarized source of $^8$Li or $^8$B is also sensitive to $j_2$ and $j_3$, and would complement $\beta$-$\alpha$ correlation measurements to allow a more precise determination of the second-forbidden terms. One measurement of the asymmetry has been performed in $^8$Li \cite{HCFH}, but was systematically skewed by $\beta$ particle scattering and required a sizable phenomenological correction. We do not include the asymmetry measurement in our analysis, but note that future measurements of this type would be helpful in constraining the values of $j_2$ and $j_3$.

Several models \cite{BARsh,BOY,KUR} have been employed to estimate the magnitude of the axial second-forbidden terms. The models predict contributions to $\delta^+$ from $j_2$ and $j_3$ which are comparable to the contributions from the induced tensor, $d_I$. It has been pointed out \cite{KDR} that mesonic exchange effects may be significant in A=8 $\beta$-decays, especially at the second-forbidden level, and that shell model calculations may break down.

To determine of the best value of $d$ from the $\delta^+$ data, second-forbidden contributions are neglected and $d$ will be assumed to take the same functional form as the Gamow-Teller matrix element, $c$. The possibility of large second-forbidden contributions to $\delta^+$, with magnitude given by the shell model predictions, will then be considered and their effect on the extracted value of $d$ will be assigned as an uncertainty. The uncertainty associated with the ambiguity in the functional form of $d$ will be estimated by fitting the $\delta^+$ data with the assumption that $d$ takes the same form as the weak magnetism operator, $b$.

\begin{figure}
\includegraphics[width=0.5\textwidth]{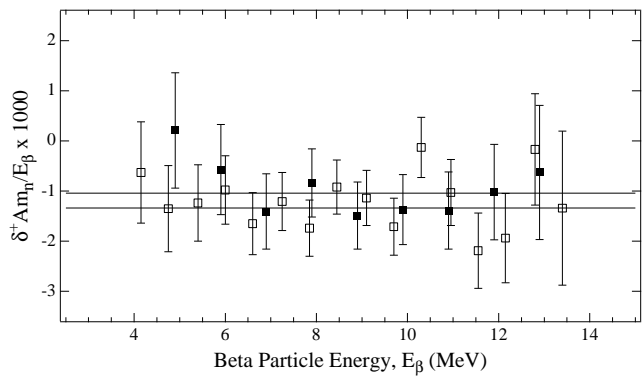}
\caption{\label{fig:deltaplus} The solid squares are experimental data on $\delta^+$ from $\beta$-$\alpha$ angular correlation measurements from Ref. \cite{TGC}, the open squares are from Ref. \cite{MGG}. The curves indicate the 1$\sigma$ error bands from the prediction for $\delta^+$ based on Eq. \ref{eq:deltaminus}, using the weak magnetism and Gamow-Teller matrix elements determined previously. Second forbidden contributions from $j_2$ and $j_3$ were ignored in this fit.}
\end{figure}

Utilizing the above assumptions, $j_2$ and $j_3$ are set equal to zero, $d$ is considered  to have the same form as $c$, $d=\eta c$, and Eq. \ref{eq:deltaplus} is used to fit the $\delta^+$ data, with $\eta$ as the only parameter. The best fit to the $\delta^+$ data from Ref. \cite{TGC} gives $\eta$=10.3(2.3) with $\chi^2$/dof=2.7/8. The best fit to the data from Ref. \cite{MGG} gives $\eta$=10.6(1.4) with $\chi^2$/dof=12.2/15. Fitting both data sets simultaneously gives $\eta$=10.5(1.2) with $\chi^2$/dof=15.0/24. The results of the fits are compared to the $\delta^+$ data in Fig. \ref{fig:deltaplus}.

The uncertainty associated with the second forbidden terms is estimated by assuming the values obtained using the model of Ref. \cite{KUR},  $j_2/A^2c\approx-$~400 and $j_3/A^2c\approx-$~750.  We take $d=\eta c$ and the $\delta^+$ data \cite{TGC,MGG} are fit, yielding $\eta$=13.8(1.2) with $\chi^2$/dof=16.5/24. 

The uncertainty associated with the unknown functional form of the induced tensor is estimated by taking $d=\xi b$. A simultaneous fit to the $\delta^+$ data sets \cite{TGC,MGG}, assuming no second forbidden contributions, gives $\xi$=0.185(20) with $\chi^2$/dof=15.3/24.

\subsection{\label{sec:NuEffects}Recoil Order Effects on the Neutrino Spectrum}

The values and uncertainties of the weak magnetism, $b$, and induced tensor, $d$, matrix elements have been deduced from experimental data. A further uncertainty is applied to these values due to imperfect isospin symmetry and electromagnetic effects. The effect of isospin breaking is estimated by comparing the Gamow-Teller matrix elements of the $^8$B and $^8$Li mirror $\beta$ decays. Previous comparisons of experimental $\alpha$ spectrum following $^8$B and $^8$Li decays indicate c$_{Li}$/c$_{B}\approx$1.07 \cite{WAR,BA}. As seen from Eqs. \ref{eq:deltaminus} and \ref{eq:deltaplus}, this uncertainty propagates linearly to the extracted values of $b$ and $d$. We thus assign to $b$(E$_x$) and $d$(E$_x$) a further 7\% uncertainty, added in quadrature with previously stated uncertainties. Further electromagnetic effects, such as the difference in decay energies of $^8$Li and $^8$B and final state electromagnetic interactions, are discussed in Ref. \cite{MGG} and are proportional to the second forbidden axial terms, $j_2$ and $j_3$. These effects contribute up to 4\%, when the largest shell model values for $j_2$ and $j_3$ are assumed. We thus add, in quadrature, a further 4\% uncertainty to $b$(E$_x$) and $d$(E$_x$).

\begin{figure}
\includegraphics[width=0.5\textwidth]{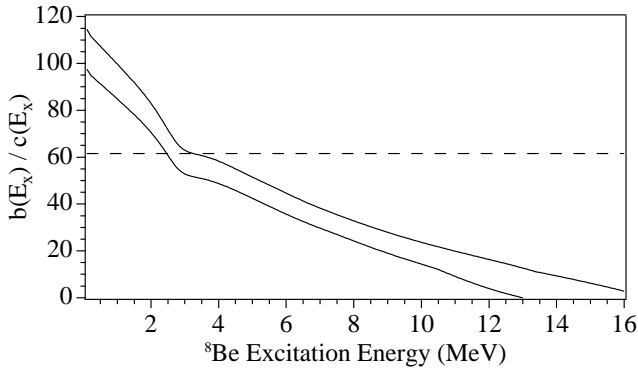}
\caption{\label{fig:wmratio} The solid curves indicate the 1$\sigma$ error bands in the ratio between the weak magnetism matrix element, $b$(E$_x$), and Gamow-Teller matrix element, $c$(E$_x$), used in this work. The dashed line represents the ratio from Ref. \cite{BH}, used in previous determinations of the neutrino spectrum \cite{BH,NFC,SNS,ORT} which neglected the excitation energy dependence of $b$(E$_x$) and $c$(E$_x$).}
\end{figure}

Fig. \ref{fig:wmratio} shows the ratio of $b$(E$_x$) to $c$(E$_x$) over the range of allowed excitation energies in $^8$Be. At high excitation energies, $c$(E$_x$) increases rapidly while $b$(E$_x$) decreases, as can be seen in Fig. \ref{fig:matrix}. In terms of the R-matrix approach, this is explained by comparing the Gamow-Teller strength of the high-lying doublet to the strength of the first excited state at 3.0~MeV, $\mathcal{M}_{A}$/$\mathcal{M}_1$=-11.8(8). For the weak magnetism transition, the ratio is much smaller, $\mathcal{M}_{A}$/$\mathcal{M}_1$=1.4(1.6) \cite{DBW}, and the doublet transition strength plays a smaller role. At excitation energies above 3.0~MeV, the result is a constructive interference of the $\mathcal{M}_{A}$ and the $\mathcal{M}_1$ terms for $c$(E$_x$). Conversely, for excitation energies below 3~MeV, the terms interfere destructively, causing $c$(E$_x$) to drop off more rapidly than $b$(E$_x$) and increasing the ratio $b$(E$_x$)/$c$(E$_x$).

The induced pseudoscalar matrix element may be estimated by applying the partially conserved axial current hypothesis, which indicates 
\begin{equation}
h(E_x)\approx\frac{4 M^2}{m_\pi^2}c(E_x).
\end{equation}
The induced pseudoscalar appears only in the last term of \ref{eq:betarecoil} which is suppressed by a factor m$_e^2$/M$^2$. The induced pseudoscalar contribution to the $\beta$ and neutrino energy spectra is thus of order m$_e^2$/m$_\pi^2$, and is ignored.

\begin{figure}
\includegraphics[width=0.5\textwidth]{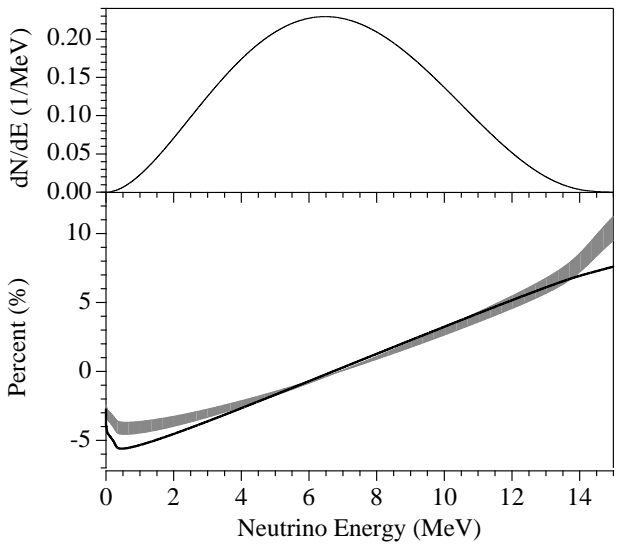}
\caption{\label{fig:recoilcompare} (Top panel) The normalized neutrino spectrum deduced in this work. (Bottom panel) The magnitude of the effect of the recoil order matrix elements on the neutrino spectrum. The grey region shows the $\pm$1$\sigma$ band of the results obtained in this work. The black line was obtained using the recommended values from Ref. \cite{BH}, which have been used in previous determinations of the neutrino spectrum.}
\end{figure}

The magnitude of recoil order effects on the $^8$B neutrino spectrum determined by the present treatment is compared to the previous treatment \cite{BH} in Fig. \ref{fig:recoilcompare}.

\section{\label{sec:Neutrino}Determination of the $^8$B Neutrino and Positron Spectra}

Radiative corrections to nuclear $\beta$ decay were first explicitly formulated in \cite{SIR}, and are exact to O($\alpha$), where $\alpha$ is the electromagnetic fine structure constant. Further corrections, dependent on the structure of the nucleus, occur at the O($\alpha^2$ln$\frac{m_n}{E_0}$) level. These model dependent corrections are insignificant when compared to the experimental uncertainties in the neutrino spectrum and are not included. Radiative corrections for the case where the neutrino is detected while the positron remains unobserved were calculated explicitly in Ref. \cite{BS}, and affect the $^8$B neutrino spectrum at the level of 1\%. 

\begin{figure}
\includegraphics[width=0.5\textwidth]{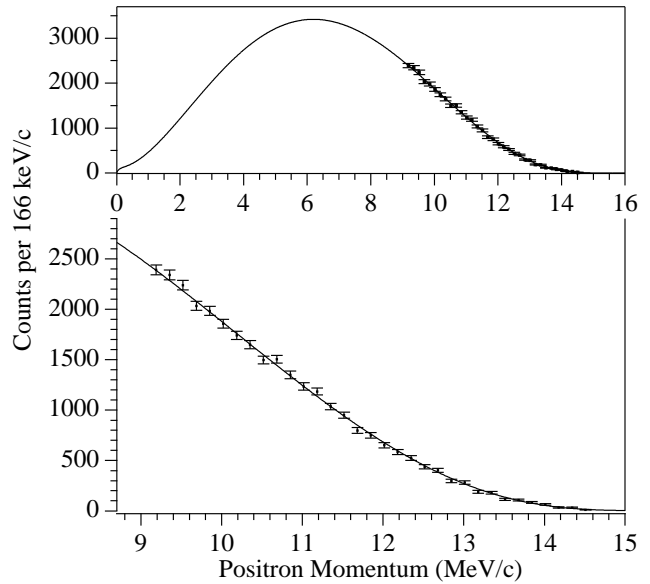}
\caption{\label{fig:positron} (Top panel) Comparison of the direct positron spectrum measurement \cite{NFC} with the predicted spectrum based on the measured alpha spectrum. The amplitude of the predicted spectrum was floated. (Bottom panel) Same comparison, showing only the momentum range measured in the direct positron spectrum measurement \cite{NFC}.}
\end{figure}

The $\beta^+$ decay strength function determined in Sec. \ref{sec:AlphaFits} was applied, using Eq. \ref{eq:betarecoil}, to determine the positron and neutrino spectra of $^8$B. 

The deduced positron spectrum was compared to the experimental spectrum \cite{NFC}, and a one parameter fit to determine the amplitude gave an agreement of $\chi^2$/dof=33.1/31, where only statistical uncertainties were included in the minimization function. The agreement is shown in Fig. \ref{fig:positron}.  The deduced positron spectrum was then allowed to float by an energy offset and marginal improvement ($\chi^2$/dof=32.6/31) was found for an offset of -14$\pm$20~keV. We note that the values quoted here are slightly different than those in Ref. \cite{WIN} due to the improved treatment of recoil order effects. The calibration uncertainty of the positron measurement is reported as 25~keV \cite{SNS}, and is not included in the fits. 

The neutrino spectrum is presented numerically, with uncertainties, in Table \ref{tab:nutable}.

\begin{table*}[!]
\caption{\label{tab:nutable}The neutrino spectrum of $^8$B and its uncertainties. Here P=dN/dE$_\nu$ is the probability of a neutrino being emitted in a given energy range. The spectrum is normalized to 1000 when integrated in terms of MeV.}
\begin{ruledtabular}
\begin{tabular}{ccccccccccccccc}
E$_\nu$ & P & $\Delta$P & E$_\nu$ & P & $\Delta$P & E$_\nu$ & P & $\Delta$P & E$_\nu$ & P & $\Delta$P & E$_\nu$ & P & $\Delta$P  \\
\hline
0.10 &  0.214 & 0.020 & 3.20 & 77.526 & 0.299 & 6.30 & 130.963 & 0.091 & 9.40 & 92.857 & -0.244 & 12.50 & 19.735 & -0.165 \\
0.20 &  0.763 & 0.043 & 3.30 & 80.456 & 0.301 & 6.40 & 131.101 & 0.074 & 9.50 & 90.528 & -0.248 & 12.60 & 17.963 & -0.157 \\ 
0.30 &  1.513 & 0.013 & 3.40 & 83.337 & 0.301 & 6.50 & 131.134 & 0.057 & 9.60 & 88.161 & -0.252 & 12.70 & 16.266 & -0.149 \\ 
0.40 &  2.507 & 0.021 & 3.50 & 86.164 & 0.302 & 6.60 & 131.063 & 0.040 & 9.70 & 85.759 & -0.255 & 12.80 & 14.647 & -0.140 \\ 
0.50 &  3.763 & 0.031 & 3.60 & 88.931 & 0.301 & 6.70 & 130.888 & 0.023 & 9.80 & 83.328 & -0.258 & 12.90 & 13.110 & -0.132 \\ 
0.60 &  5.239 & 0.041 & 3.70 & 91.635 & 0.300 & 6.80 & 130.611 & 0.007 & 9.90 & 80.869 & -0.261 & 13.00 & 11.655 & -0.123  \\ 
0.70 &  6.914 & 0.053 & 3.80 & 94.272 & 0.298 & 6.90 & 130.232 & -0.010 & 10.00 & 78.387 & -0.263 & 13.10 & 10.286 & -0.115  \\ 
0.80 &  8.772 & 0.065 & 3.90 & 96.839 & 0.296 & 7.00 & 129.752 & -0.027 & 10.10 & 75.885 & -0.264 & 13.20 &  9.005 & -0.106  \\ 
0.90 & 10.798 & 0.077 & 4.00 & 99.331 & 0.292 & 7.10 & 129.174 & -0.039 & 10.20 & 73.368 & -0.265 & 13.30 &  7.813 & -0.097  \\ 
1.00 & 12.976 & 0.091 & 4.10 & 101.746 & 0.288 & 7.20 & 128.497 & -0.051 & 10.30 & 70.837 & -0.265 & 13.40 &  6.712 & -0.088  \\ 
1.10 & 15.292 & 0.104 & 4.20 & 104.081 & 0.284 & 7.30 & 127.724 & -0.063 & 10.40 & 68.298 & -0.265 & 13.50 &  5.703 & -0.080  \\ 
1.20 & 17.735 & 0.118 & 4.30 & 106.332 & 0.278 & 7.40 & 126.856 & -0.075 & 10.50 & 65.754 & -0.264 & 13.60 &  4.787 & -0.071  \\ 
1.30 & 20.292 & 0.132 & 4.40 & 108.497 & 0.273 & 7.50 & 125.895 & -0.087 & 10.60 & 63.209 & -0.263 & 13.70 &  3.965 & -0.062 \\ 
1.40 & 22.950 & 0.145 & 4.50 & 110.574 & 0.266 & 7.60 & 124.843 & -0.097 & 10.70 & 60.667 & -0.262 & 13.80 &  3.237 & -0.054  \\ 
1.50 & 25.699 & 0.158 & 4.60 & 112.560 & 0.259 & 7.70 & 123.701 & -0.108 & 10.80 & 58.131 & -0.259 & 13.90 &  2.602 & -0.046  \\ 
1.60 & 28.528 & 0.172 & 4.70 & 114.452 & 0.252 & 7.80 & 122.471 & -0.118 & 10.90 & 55.606 & -0.257 & 14.00 &  2.058 & -0.038  \\ 
1.70 & 31.427 & 0.184 & 4.80 & 116.250 & 0.244 & 7.90 & 121.156 & -0.128 & 11.00 & 53.095 & -0.254 & 14.10 &  1.602 & -0.031  \\ 
1.80 & 34.386 & 0.197 & 4.90 & 117.951 & 0.236 & 8.00 & 119.758 & -0.138 & 11.10 & 50.602 & -0.251 & 14.20 &  1.228 & -0.024  \\ 
1.90 & 37.395 & 0.208 & 5.00 & 119.553 & 0.227 & 8.10 & 118.278 & -0.148 & 11.20 & 48.131 & -0.247 & 14.30 &  0.929 & -0.019  \\ 
2.00 & 40.447 & 0.219 & 5.10 & 121.056 & 0.218 & 8.20 & 116.720 & -0.158 & 11.30 & 45.686 & -0.242 & 14.40 &  0.694 & -0.014  \\ 
2.10 & 43.531 & 0.230 & 5.20 & 122.457 & 0.209 & 8.30 & 115.086 & -0.166 & 11.40 & 43.271 & -0.238 & 14.50 &  0.513 & -0.011 \\ 
2.20 & 46.640 & 0.240 & 5.30 & 123.755 & 0.199 & 8.40 & 113.378 & -0.175 & 11.50 & 40.889 & -0.233 & 14.60 &  0.376 & -0.008  \\
2.30 & 49.767 & 0.249 & 5.40 & 124.951 & 0.189 & 8.50 & 111.599 & -0.185 & 11.60 & 38.545 & -0.227 & 14.70 &  0.273 & -0.006  \\
2.40 & 52.903 & 0.258 & 5.50 & 126.042 & 0.179 & 8.60 & 109.751 & -0.193 & 11.70 & 36.242 & -0.222 & 14.80 &  0.196 & -0.004  \\
2.50 & 56.041 & 0.266 & 5.60 & 127.028 & 0.168 & 8.70 & 107.838 & -0.200 & 11.80 & 33.984 & -0.215 & 14.90 &  0.140 & -0.003  \\
2.60 & 59.174 & 0.273 & 5.70 & 127.909 & 0.157 & 8.80 & 105.862 & -0.208 & 11.90 & 31.774 & -0.209 & 15.00 &  0.099 & -0.002  \\
2.70 & 62.296 & 0.279 & 5.80 & 128.683 & 0.146 & 8.90 & 103.827 & -0.215 & 12.00 & 29.616 & -0.202 & 15.10 &  0.069 & -0.002  \\
2.80 & 65.401 & 0.285 & 5.90 & 129.351 & 0.135 & 9.00 & 101.734 & -0.222 & 12.10 & 27.515 & -0.195 & 15.20 &  0.047 & -0.001  \\ 
2.90 & 68.482 & 0.289 & 6.00 & 129.914 & 0.124 & 9.10 & 99.587 & -0.228 & 12.20 & 25.472 & -0.188 & 15.30 &  0.000 & 0.000  \\
3.00 & 71.533 & 0.293 & 6.10 & 130.369 & 0.113 & 9.20 & 97.390 & -0.234 & 12.30 & 23.493 & -0.181 & 15.40 &  0.000 & 0.000  \\
3.10 & 74.549 & 0.296 & 6.20 & 130.719 & 0.102 & 9.30 & 95.146 & -0.239 & 12.40 & 21.579 & -0.173 & 15.50 &  0.000 & 0.000  \\
\end{tabular}
\end{ruledtabular}
\end{table*}

\section{\label{sec:Conclusions}Conclusions}

An accurate determination of the $^8$B neutrino spectrum is important for the proper analysis of solar neutrino data. Measurements of the $\alpha$ energy spectrum following the $^8$B $\beta^+$ decay provide the most direct method of inferring the $\beta^+$ decay strength function, which is used to predict the neutrino spectrum. The $\alpha$ spectrum experiment described in this work was designed to eliminate several of the systematic effects common to past measurements, and precisely determines the strength function characterized by the many level R-matrix approximation.

The primary uncertainty in the $\alpha$ spectrum measurement arises from the determination of energy scale, which was dominated by the temporal gain shift over the seven day run. Uncertainties in calibration from the implanted $^{20}$Na source and the correction for positron energy loss in the detector also contribute. Uncertainty in the energy scale near the spectrum peak was 9~keV. The uncertainties in the $\alpha$ spectrum were propagated to the neutrino spectrum using the R-matrix approach. 

Recoil order effects provide a significant contribution to the $^8$B $\beta$ decay, and have been treated in this work using the best available experimental data. The results differ from the previous treatment \cite{BH} which has been applied in recent determinations of the neutrino spectrum \cite{BH,NFC,SNS,ORT}. 

The primary component of the recoil order corrections is the weak magnetism term, $b$. Uncertainties in $b$ are due to experimental effects \cite{DBW} as well as isospin breaking and electromagnetic effects, and are included as uncertainties in the neutrino spectrum. Uncertainties in the neutrino spectrum from recoil order effects are roughly half as large as the uncertainties from the $\alpha$ spectrum measurement.

The $\alpha$ spectrum reported here is in substantial disagreement with the previous measurement of comparable precision \cite{ORT}, but is in good agreement with the direct measurement of the positron spectrum \cite{NFC}.

This work was supported by the Department of Energy under Contract Nos. W-31-109-ENG-38 and DE-AC03-76SF00098.
 
\bibliography{PRC.bib}

\end{document}